\documentclass{iaus}
\usepackage{graphicx}

\title[The Exoplanet Albedo] 
{Towards the Albedo of an Exoplanet: \\
MOST Satellite Observations of \\
Bright Transiting Exoplanetary Systems}

\author[Rowe et al.]   
{Jason F. Rowe,$^1$ Jaymie M. Matthews, Sara Seager, \\ 
Dimitar Sasselov, Rainer Kuschnig,  David B. Guenther,\\ 
Anthony F.J. Moffat, Slavek M. Rucinski, \\
Gordon A.H. Walker and Werner W. Weiss}

\affiliation{$^1$NASA-Ames Research Park\\
MS-244-30\\
Moffett Field, CA 94035-1000\\
email: {\tt jasonfrowe@gmail.com}}

\pubyear{2008}
\volume{253}  
\pagerange{1--7}
\jname{Transiting Planets}
\editors{D. Sittler, D. Keon, D. Gilmour}
\begin{document}

\maketitle

\begin{abstract}
The Canadian \textit{MOST} satellite is a unique platform for observations of bright transiting exoplanetary systems.  Providing nearly continuous photometric observations for up to 8 weeks, \textit{MOST} can produce important observational data to help us learn about the properties of exosolar planets.  We review our current observations of HD 209458, HD 189733 with implications towards the albedo and our progress towards detecting reflected light from an exoplanet.

\end{abstract}

\firstsection 
\section{Introduction}

We train our detectors to monitor the periodic cycle of a planet transiting a star, being eclipsed by the star and the change of the planet's illuminated phase.  Our primary goal: to determine what fraction of photons come from the star and what fraction come from the planet.  The easiest feature to detect and measure is when the planet transits the star from the observers perspective where the depth of the transit is determined by the ratio of the projected surface area of a planet and star.\footnote{Ignoring emission and transmission of photons from the planet and assuming uniformity on the surface of the star.}  When a planet is eclipsed by the star only photons emitted by the star are detected.  The ratio of the observed flux before first contact to the observed flux during total eclipse determines the ratio of the planet's reflected flux to the star's incident flux at the substellar point and determine the geometric albedo.

With the \textit{MOST} ({\it Microvariablity and Oscillations of STars}) satellite we have observed two exoplanetary systems with a goal of measuring the reflected light signature from the planet.  For our first candidate, HD 209458 (G0V, V=7.7), observations spanning 14, 45 and 29 days have been obtained in 2004, 2005 and 2007 respectively.  From the 2004 and 2005 observation runs we set a $1\, \sigma$ upper limit on the planet-to-star flux ratio of 16 ppm.  Details of this study may be found in \cite[Rowe \etal\ (2006)]{row06} and \cite[Rowe \etal\ (2007)]{row07}.  We have also observed the chromospherically active star HD 189733 (K2V, V=7.7) in an attempt to measure the albedo of the transiting Jupiter-mass planet.  Our observations span 22 and 31 days for our 2006 and 2007 campaigns.  The presence of large spots and large scatter seen in optical photometry has hindered our progress.

\subsection{The Albedo}

Understanding what an albedo tells us about the nature of a planet requires a quick review of the definition of the Bond albedo, which in turn defines the geometric albedo and the phase function.  Bond's definition of an albedo, from 1861, is: ``Let a sphere S be exposed to parallel light.  Then its albedo ($A_B$) is the ratio of the whole amount of light reflected from S to the whole amount incident on it" \cite[(Russell 1916)]{rus16}.  The incident flux ($L_{in}$) is the stellar flux density at the distance of the planet from the star ($a$) times the cross section of the planet ($\pi R_p^2$) which, by definition of the Bond albedo states, that the emergent flux is $L_{out}=A_B\ L_{in}$.  For an anisotropic scatter the planetary flux density observed at a distance $D$ is 
\begin{equation}\label{eq:Fp}
F_p=C\ \phi (\alpha)\ \frac{L_{out}}{4 \pi D^2},
\end{equation}
where $C$ is a constant and $\phi$ is the phase function describing the brightness of the planet at phase angle $\alpha$ normalized such that $\phi(\alpha = 0) = 1$.  Integrating over the surface of a sphere with radius $D$,
\begin{equation}
L_{out} = \int_S C\ \phi (\alpha)\ \frac{L_{out}}{4 \pi D^2}\ dS,
\end{equation}
we can solve for the normalization constant $C$ as the flux density integrated over the surface of the planet much equal total flux received.  We find,
\begin{equation}\label{eq:C}
C=\frac{2}{\int_0^{\pi} \phi (\alpha)\ {\rm sin} \alpha\ d\alpha} = \frac{4}{q},
\end{equation}
where we have defined the phase integral,
\begin{equation}
q= 2 \int_0^{\pi} \phi(\alpha)\ {\rm sin} \alpha\ d\alpha,
\end{equation}
which quantifies the scattering properties of the reflective surface over all phase angles and is proportional to the total amount of light reflected by the planet.
The geometrical albedo is defined at phase $\alpha=0$.  Using the definition of the Bond albedo with Equation \ref{eq:Fp} and replacing $L_{in}$ with the flux density of the star ($F_*$) at the planet-to-star distance ($a$) we obtain,
\begin{equation}\label{eq:fp}
F_p = C\ A_B \frac{F_* R_p^2}{4 a^2},
\end{equation} 
which we rearrange to solve for the Bond albedo and substitute Equation \ref{eq:C} for $C$ and obtain,
\begin{equation}
A_B = \frac{4}{C} \frac{F_p}{F_*} \frac{a^2}{R_p^2} = \frac{F_p}{F_*} \frac{a^2}{R_p^2} q = A_g\ q,
\end{equation}
which gives us the geometric albedo ($A_g$).  The geometric albedo can be physically interpreted as the ratio of the planetary flux density at full phase ($\alpha=0$) to a Lambertian surface of the same cross section.  The ratio of the planetary to stellar flux can be written as an approximation given by,
\begin{equation}\label{ag_approx}
\frac{F_p}{F_*} \approx A_g \left( \frac{R_p}{R_J} \right)^2 \left( \frac{a}{0.05\ {\rm AU}} \right)^{-2}  100\ \rm{ppm},
\end{equation}
where $R_J$ is the radius of Jupiter.  Equation \ref{ag_approx} demonstrates that ideal exoplanet candidates for measuring low amplitude reflected light signatures have large radii and orbit at small distances from its host star. For close-in planets with radii similar to Jupiter and have a geometric albedo of 10\% the photometric signature of planet being eclipsed by the star has an amplitude of approximately 10 ppm.

\section{Measurements and Upper Limits}

\begin{figure}[b]
\begin{center}
 \includegraphics[width=4.5in]{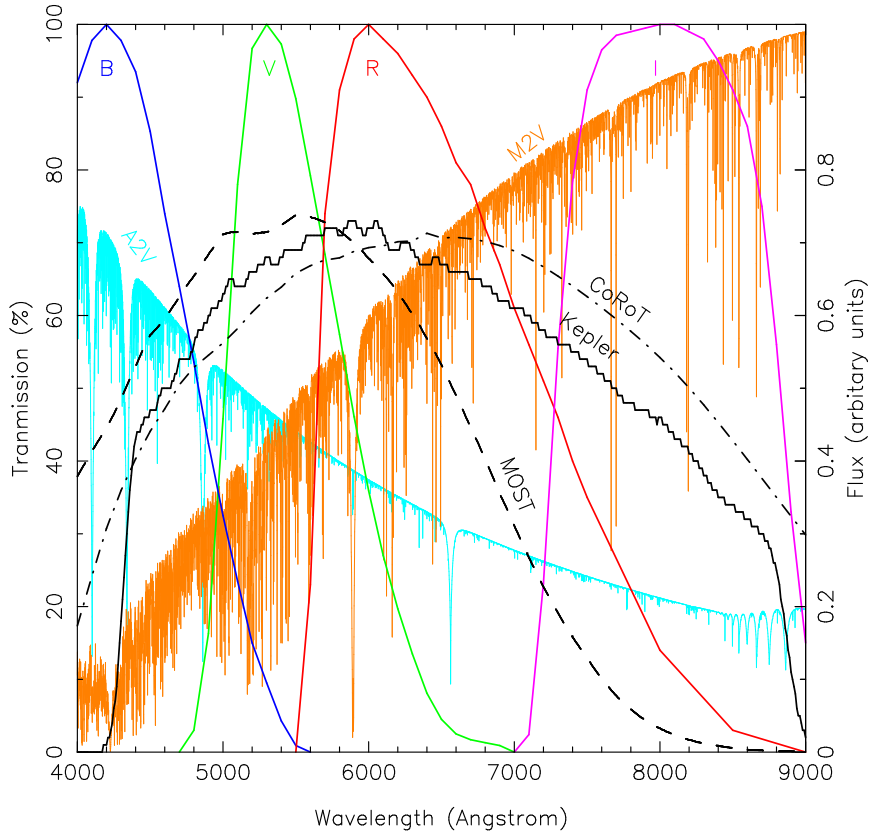} 
 \caption{The transmission functions for the Johnson B,V,R,I filters are shown from left to right in blue, green, red and magenta respectively and have been scaled to peak at 100\% transmission. The \textit{MOST} bandpass bandpass is marked by the dashed line, the Kepler bandpass is shown in black and the CoRoT whitelight bandpass is shown by the dot-dashed line.  The spectrum for an A2V star is shown in cyan, which peaks in the UV and the spectrum for a M2V star is shown in orange which peaks in the infrared.  The two spectra have been scaled to have equal flux in the Johnson V filter.}
   \label{fig1}
\end{center}
\end{figure}

The \textit{MOST} satellite \cite[(Walker, Matthews \etal\ 2003; Matthews \etal\ 2004)]{wal03,mat04} houses a 15-cm optical telescope feeding a CCD photometer through a single custom broadband (400 - 700 nm) filter, as shown in Figure \ref{fig1}.  From its 820-km-high polar Sun-synchronous orbit, \textit{MOST} can monitor stars in its equatorial Continuous Viewing Zone for up to 8 weeks without interruption.  Classified as a microsatellite (mass = 54 kg; peak solar power = 39 W) \textit{MOST} has limited onboard processing capability, memory, and downlink. Hence it is not possible to transfer the entire set of 1024 $\times$ 1024 pixels of the Science CCD to Earth at a rapid sampling rate and with an ADC (analogue-to-digital Conversion) of 14 bits (necessary to preserve variability information at the ppm level). Small segments of the CCD (”subrasters”) are stored, which contain key portions of the target field. This usually includes any combination of a Primary Science Target Fabry Image and up to 6 Direct Imaging targets depending on the image acquisition rate.  

Direct Imaging consists of defocused star images (FWHM $\sim$ 2.5 pixels) that are projected onto an open area of the Science CCD adjacent to the Fabry lens array.  Subrasters of typical dimensions 20 $\times$ 20 pixels are placed around each Direct Image target It is possible to select a star as bright as V = 6.5 as the principal science target in the Direct Imaging field. Then the exposure time and sampling rate can be optimized for the Direct Imaging target. This was the case for HD 209458 and HD 189733.  Details of the data reduction steps can be found in \cite[Rowe \etal\ (2006)]{row06} and \cite[Rowe \etal\ (2007)]{row07}.

\begin{figure}[b]
\begin{center}
\includegraphics[width=5.2in]{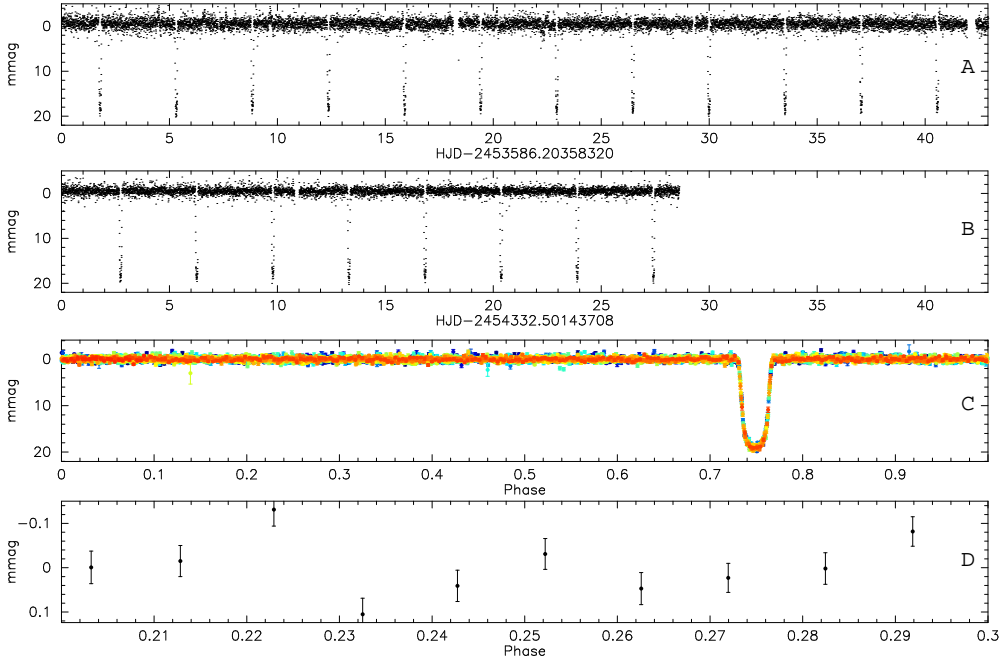}
 \caption{Observations of HD 209458: Panel A shows the lightcurve for the 2005 observations using 5 minute bins.  Panel B shows the lightcurve of the the 2007 observations using 5 minute bins.  Panel C shows the combined 2004, 2005 and 2007 lightcurve phased to the orbital period of HD 209458b.  Panel D shows a portion of the phased lightcurve heavily binned with $1\, \sigma$ errorbars centered around the expected occurrence of the planet being eclipsed by the star.}
   \label{fig:hd209458}
\end{center}
\end{figure}

\begin{figure}[b]
\begin{center}
 \includegraphics[width=5.2in]{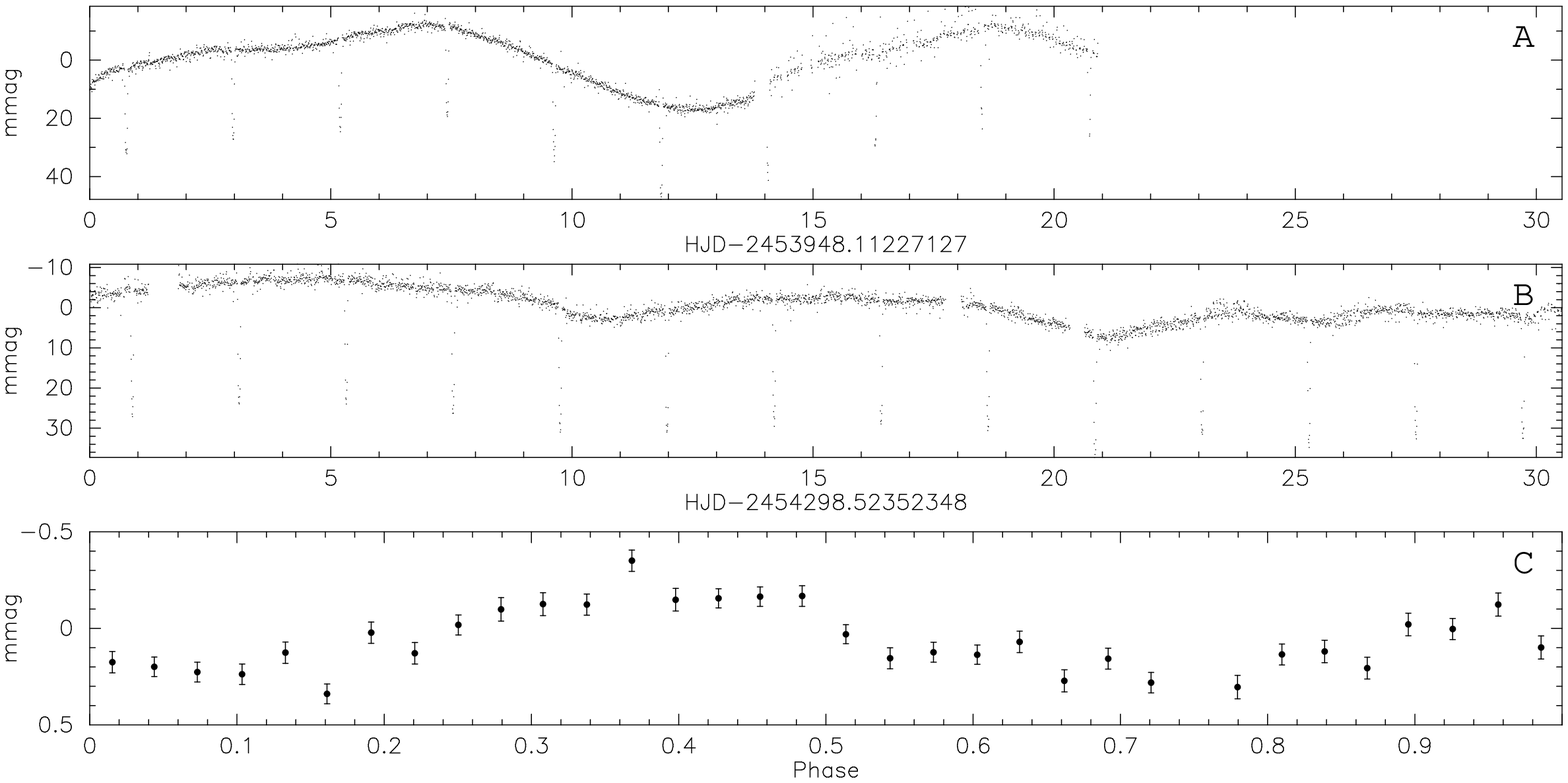} 
 \caption{Observations of HD189733:  Panel A shows the 2006 observations binned with 10 minute intervals for HD 189733.  Panel B shows the 2007 observations of HD 189733 binned with 10 minute intervals.  Panel C shows the phase lightcurve after the spot modulation in the lightcurve has been filtered out.  The data have been binned at 0.03 in phase.  The eclipse occurs at phase 0.25.  The large amplitude primary transit occurs at phase 0.75 but is not visible due to small vertical scale chosen to display variations with phase.}
   \label{fig:hd189733}
\end{center}
\end{figure}

\subsection{Upper Limits on the Albedo (HD 209458b)}\label{hd209458}

The planetary transit lasts for $\sim$3.7 hours (and hence the eclipse).  To measure the secondary eclipse for the planet disappearing behind the star, we average the photometry over phase bins of width 0.044 ($\sim$3.7 hours) centred 0.5 phase away from the transit (at phase 0.25 in Panel C of Figure \ref{fig:hd209458}).  We then find the average of the photometry from two adjacent bins with the same width.  Our error is estimated by bootstrapping the means for each bin.  Using our 2004 and 2005 datasets this basic approach gives a flux ratio of the star and planet of $22\pm29$ ppm. 

If we make the assumption that the planet scatters as a Lambert sphere then $\phi(\alpha)$ is simply a function of the phase of the planet from the observer's perspective,
\begin{equation}
\phi(\alpha)=\frac{1}{\pi} ( {\rm sin} \alpha + (\pi - \alpha) {\rm cos} \alpha ),\  q=\frac{3}{2}.
\end{equation}
Using Equation \ref{eq:fp} and allowing for an arbitrary phase angle, the bolometric light curve, as a function of phase ($\alpha$) is,
\begin{equation}\label{eq:ag}
\frac{F_p}{F_*} = \frac{2}{3} A_B \frac{R_p^2}{a^2} \phi(\alpha).
\end{equation}
Our observations are not bolometric, thus we make the assumption that
\begin{equation}
A_{MOST} \simeq A_g \simeq \frac{2}{3}A_B,
\end{equation}
where $A_{MOST}$ is the geometric albedo measured over the \textit{MOST} bandpass.  For Sun-like host stars, the peak of the emitted spectral energy distribution is in the optical range, thus the \textit{MOST} bandpass geometric albedo is a good approximation of the geometric albedo.  Our second assumption is that the planet scatters light as a Lambert sphere.  The Lambert sphere is a reflecting surface with a reflection coefficient that is constant for all angles of incidence.  The reflection coefficient is the ratio of the amount of light diffusively reflected in all directions by an element of the surface to the incident amount of light that strikes that element (see \cite[Rowe \etal\ (2006)]{row06} for further details).  Our best fit to the flux ratio of the planet and star ($F_p/F_*$) is 7 ppm with a $1\sigma$ upper limit of 16 ppm which gives the geometric albedo measured through the \textit{MOST} filter: $A_{\rm MOST} = 3.8 \pm 4.5\%$.

Panels A and B in Figure \ref{fig:hd209458} shows our 2005 and 2007 photometric measurements using 5 minute bins.  The 2004 and 2005 observations used 1.5s exposures, and 1 exposure was obtained every 10s to maintain a data acquisition rate compatible with downlink capabilities.  During the 2006 observing season an additional capability was added to the satellite onboard software that allows image stacking.  The 2007 HD 209458 observations consist of one co-added exposure sampled every 22.5 seconds. A single co-added exposure constitutes 15 stacked 1.5s exposures.  With less dead time between exposures and the resulting increase in the number of detected photons the extracted photometry shows less scatter.  This can be seen from a comparison of Panels A and B in Figure \ref{fig:hd209458}.  Both lightcurves have been plotted on the same scale for comparison.  Our next task will be to repeat our measurements of the albedo using our 2007 observations.

\subsection{Upper Limits on the Albedo (HD 189733b)}\label{hd189733}

HD 189733 was observed by \textit{MOST} for 21 days in 2006 and 31 days in 2007.  Individual exposures lasted 1.5 s, with 14 consecutive images stacked on board the satellite to produce one co-added exposure sampled every 21 seconds. Examination of Panel A of Figure \ref{fig:hd189733} shows a noticeable change in the quality of the photometry for the last third of the data. The first 14 days of observation have a duty cycle of 94\% and the final 7 days have a reduced duty cycle of 46\%.  The last week of observations was shared with another \textit{MOST} Primary Science Target field.   When the intrinsic variability of the star was recognized in the early photometry, the originally planned 14-day run was extended by a week, by switching with another target during every \textit{MOST} satellite orbit ($P = 101.4$ min).  This meant that observations of HD 189733 could only be carried out during phases of the highest scattered Earthshine, resulting in increased scatter in the light curve which can be treated in the reduction.  The 2007 observations were uninterrupted.

The obvious feature in the lightcurves is rotational modulation due to starspots.  The amplitude was approximately 3\% in 2006 and 1\% in 2007.  Modeling the reflected light signature requires removal of the star spot influence on the light curve. The spot modulation was filtered out by prewhitening all significant frequencies not related to the planetary orbit or its harmonics.  The Fourier Transform (FT) of a lightcurve with a periodic transit will show the primary period of the transit and its harmonics with a comb-like pattern.

Visual inspection of Panel C in Figure \ref{fig:hd189733} appears to indicates the presence of a reflected light signature at high significance.  Infrared detection of the planetary eclipse shows that the eclipse should occur at phase of 0.25 in Figure \ref{fig:hd209458}.  Thus we should see a drop in the brightness of the system equal to the amplitude of the phase variations (or even larger if the atmospheric properties of the planet produce strong backscatter).  No such variation is detected.  The likely explanation is that stellar activity cannot be fully removed by our filting method.  Using a FT to remove significant periodicities in the data, requires that the stellar activity is periodic.  Long lifetime starspots produce this behaviour from stellar rotation, but flare activity and short lived starspots will not produce this kind of behaviour and can remain in our filtered light curve.  Thus active stars seriously inhibit our ability to measure reflected light from the planet.  With continued observations of HD 189733, one can hope that the stellar activity phased to the orbital period of the planet will average out leaving behind a signature of a reflected light.

\section{Discussion and the Future}

In the absence of clouds, all hot Jupiter models do predict extremely low visible-wavelength geometric albedos, due to strong, broad absorption lines of neutral atomic Na and K.  Some model atmospheres include self-consistent cloud formation in a 1D, complete cloud cover scenario (\cite[Ackerman \& Marley 2001]{ack01}, \cite[Cooper \etal\ 2003]{coo03}). Nevertheless, the physics of cloud formation is a process that is not well understood or constrained both for particle size and densities which are governed by the competing effects of condensation and coagulation versus sedimentation (Marley et al. 1999). The low reflected fraction of incident radiation, as measured for HD 209458b, readily rules out reflective clouds.  New Spitzer observations at 3.6, 4.5, 5.8 and $8.0\, \mu$m \cite[(Knutson \etal\ 2007)]{knu07} indicate a temperature inversion which requires an extra, unknown absorber at low pressures \cite[(Burrows \etal\ 2007)]{bur07}.  Our low albedo limit means that, if the absorber is a cloud, then it must not be highly reflective.

With the 2007 photometric reductions for the HD 209458 and HD 189733 datasets nearing completion, we plan to refine our albedo estimates for both planets.  With the addition of onboard image stacking we gain a factor better than 6 for number of detected photons and the new datasets have a potential to push the albedo limits to a few percent at high significance. Both HD 189733 and HD 209458 are scheduled to be reobserved in 2008.

\textit{MOST} will also observe the highly eccentric (e=0.93) planetary system HD 80606 (G5V, V=8.9).  The large eccentricity means that the planet distance ranges from 0.03 to 0.84 AU over a 111 day period.  The large eccentricity also means that the dayside temperature of the planet is expected to change from $\sim$ 750 K to 1250 K on the timescale of a week.  This offers the potential to study the reflective properties of an exoplanet as a function of temperature.

Rayleigh scattering, Mie scattering and molecular absorption are the dominant mechanisms that determine the reflected and emitted spectra of an extrasolar giant planet \cite[(Marley \etal\ 2006)]{mar06}.  In the bluer portion of the optical spectra (wavelengths shorter than 600nm) Rayleigh scattering in a clear atmosphere will reflect a large fraction of the stellar flux outwards. In the red portion (wavelengths greater than 600nm) photons will be absorbed deep in the atmosphere that will make the reflected spectrum relatively dark.  Towards the infrared, the spectrum will begin to be dominated by thermal radiation from the planet. Figure \ref{fig1} shows the bandpass for the \textit{Kepler} and \textit{CoRoT} (whitelight) spacecraft.  Since the bandpasses of these missions, including \textit{MOST}, extends towards the red (and beyond) portion of the spectrum, the reflected light signature for planets that orbit close to the host stars can be contaminated by thermal radiation.  As a final note, when attempting to measure low albedos around giant close-in exoplanets precaution should be taken ensure that reflected light signatures are not confused by similar lightcurves produced by thermal radiation.

\end{document}